\documentclass[final,3p,times,twocolumn,floatfix]{elsarticle}

\usepackage{epsfig}

\usepackage{amssymb}

\journal{Computational Materials Science}

\newcommand{\dimtemp}{\Delta}

\begin{document}

\begin{frontmatter}



\title{From wetting to melting along grain boundaries using phase field and sharp interface methods}


\author[mpie]{V. Sai Pavan Kumar Bhogireddy}
\author[mpie]{C. H\"uter}
\author[mpie]{J. Neugebauer}
\author[icams]{O. Shchyglo}
\author[icams]{I. Steinbach}
\author[mpie]{R. Spatschek}

\address[mpie]{Max-Planck-Institut f\"ur Eisenforschung GmbH, 40237 D\"usseldorf, Germany}
\address[icams]{Interdisciplinary Centre for Advanced Materials Simulation, Ruhr-Universit\"at Bochum, 44780 Bochum, Germany}

\begin{abstract}
We investigate the ability of a multi-order parameter phase field model with obstacle potentials to describe grain boundary premelting in equilibrium situations.
In agreement with an energetic picture we find that the transition between dry and wet grain boundaries at the bulk melting point is given by the threshold $2\sigma_{sl}=\sigma_{gb}$, with $\sigma_{sl}$ being the solid-melt interfacial energy and $\sigma_{gb}$ the energy of a dry grain boundary.
The predictions for premelting are confirmed by simulations using the phase field package {\sc OpenPhase}.
For the prediction of the kinetics of melting along grain boundaries in pure materials, taking into account the short ranged interactions which are responsible for the grain boundary premelting, a sharp interface theory is developed.
It confirms that for overheated grain boundaries the melting velocity is reduced (increased) for non-wetting (wetting) grain boundaries.
Numerical steady state predictions are in agreement with a fully analytical solution in a subset of the parameter space.
Phase field simulations confirm the predictions of the sharp interface theory.

\end{abstract}


\end{frontmatter}


\section{Introduction}
\label{intro::section}

Grain boundaries are naturally present in many materials, and their understanding is essential for improving their properties.
In particular at higher temperatures grain boundary induced failure can occur, for example during hot cracking \cite{Rappaz:2003aa}. 
Here in particular grain boundary melting  \cite{Mishin:2009aa} can play a significant role, caused by the overlap of adjacent solid-melt interfaces, which lead to an effective interaction between them, expressed through the so called disjoining potential.
This effect has been studied experimentally \cite{Glicksman:1972aa, Hsieh19891637, Alsayed:2005aa}, theoretically \cite{Widom:1978aa,Rappaz:2003aa}, and computationally, using lattice models \cite{Besold:1994aa,PhysRevB.21.1893}, molecular dynamics or Monte Carlo simulations \cite{PhysRevE.79.020601, Williams20093786}, phase field models \cite{PhysRevE.81.051601}, orientational order parameter phase field models \cite{Tang:2006aa,Lobkovsky2002202}, phase field crystal \cite{PhysRevB.78.184110,Adland:2013ys,PhysRevB.77.224114,Olmsted:2011vn} and amplitude equations descriptions \cite{Spatschek:2010fk,Huter:2014aa,kar13}.
In general, high angle grain boundaries tend to premelt, i.e.~a thin melt layer can appear along the grain boundary already below the bulk melting point.
Energetically, the ratio of the dry grain boundary energy $\sigma_{gb}$ to twice the value of the solid-melt interfacial energy $2\sigma_{sl}$ is the relevant parameter at the melting temperature $T_M$:
For $\sigma_{gb}/2\sigma_{sl}$ larger (smaller) than one the dry grain boundary is energetically less (more) favourable;
hence one expects a repulsive (attractive) interaction between the solid-melt interfaces.

In this article we investigate how grain boundary premelting appears in phase field models, which are frequently used for various aspects of microstructure evolution \cite{KarmaReview, ChenReview, steinbach, Steinbach:2013aa, Spatschek11}.
The multi-order parameter phase field model \cite{Steinbach:1998aa} is often used for the simulation of such problems.
It is the basis of the phase field codes {\sc Micress} \cite{EikenJ2006} and {\sc OpenPhase} \cite{openphase}.
In a phase-field context, interactions between solid-melt interfaces appear when the smooth order parameter profiles with a width $\xi$ overlap.
Despite the similarity to other models with a double well potential \cite{Folch:2005kx}, the premelting behavior is different here \cite{PhysRevE.81.051601,Bhogireddy:2014aa}.
Next to a theoretical analysis we perform here numerical simulations to validate the predictions of the short ranged interactions and grain boundary premelting.

Beyond equilibrium situations the kinetics of grain boundary premelting is of highest interest.
Recently, the heterogeneous nucleation of liquid droplets at overheated grain boundaries has been studied using atomistic and continuum methods \cite{Frolov:2011fk}.
This way, a framework has been developed to incorporate the aforementioned short ranged interactions to nucleation processes.
The subsequent growth regime in the diffusion limited has been studied in \cite{Huter:2014uq} using sharp interface methods.
In contrast to \cite{EAB_CH_DP_DET_PRL99_105701_2007} also the effect of the disjoining potential has been considered there.
A central outcome is that it has --- although relevant only in the triple junction region formed by the advancing melt front with the grain boundary --- a quite substantial influence on the melting velocity.
In the present article we aim additionally at a complementary modelling using phase field descriptions.

The article is organised as follows:
In section \ref{model::section} we introduce the multi-order parameter phase field model, which is the basis for the theoretical and numerical investigations of grain boundary premelting and the wetting kinetics in this article.
Section \ref{wettingsection} is devoted to static equilibria.
The analytical prediction of the premelting of planar grain boundaries is presented is section \ref{multi:section}.
A complementary numerical investigation using the {\sc OpenPhase} code is made in section \ref{OPFGBpremelting}.
The kinetics of the wetting of a low angle grain boundary is discussed in section \ref{kinetics::section}.
The simulation of this phenomenon by phase field simulations in section \ref{PFmelting::subsection} is followed by a sharp interface analysis in section \ref{sharpinterfacemelting::subsection}, including a comparison of the two perspectives.
The main results are summarised in section \ref{summary}.

\section{The multi order parameter phase field model}
\label{model::section}

Since the multi-order parameter phase field model by Steinbach and Pezzolla \cite{Steinbach:1998aa} plays a central role for the analysis and simulations in this article, we concisely summarise the governing equations here.
In the basic situation of a single component, multiphase or polycrystalline structure the model is described by the free energy
\begin{eqnarray}
 F_0 &=& \int \Bigg\{ \sum_{\alpha=1}^N \sum_{\beta >\alpha}^N \Big(\frac{-4\eta_{\alpha\beta}\sigma_{\alpha\beta}}{\pi^2}\nabla\phi_\alpha \nabla\phi_\beta \nonumber \\
&+& \frac{4\sigma_{\alpha\beta}}{\eta_{\alpha\beta}}\phi_\alpha \phi_\beta \Big) \nonumber \\
&-&  \frac{L(T-T_M)}{T_M}[1-g(\{\phi_\alpha\})] \Bigg\} dV.  \label{intro::eq1}                                           
\end{eqnarray}
The dimensionless phase fields (order parameters) $\phi_\alpha$, which vary between $0$ and $1$, distinguish between the phases or grains.
In the summations, $N$ is the maximum number of phases which may appear in the description.
The constraint
\begin{equation}
\sum_\alpha^N \phi_\alpha=1, \label{intro::eq2}
\end{equation}
which has to hold everywhere, allows to interpret the phase fields as local volume fractions of the phases.
In contrast to models with a double- or multi-well potential, the confinement of the order parameters to the interval $\phi_\alpha\in[0:1]$ is enforced by an infinite energy penalty if the phase field values are outside this domain.
This is formally described by an additional energy term
\begin{equation} \label{intro::eq3}
F = F_0 + \left\{ 
\begin{array}{cc}
0 & 0\leq \phi_\alpha\leq 1, \alpha=1, \ldots, N \\
\infty & \mbox{else}
\end{array}
\right. .
\end{equation}
This so called multi-obstacle potential strictly confines the smoothing of the phase fields at an interface to a finite layer.
The other parameters in Eq.~(\ref{intro::eq1}) are the interfacial energies $\sigma_{\alpha\beta}=\sigma_{\beta\alpha}$ (dimension: energy/area) and the interface thicknesses $\eta_{\alpha\beta}=\eta_{\beta\alpha}$ (dimension: length).
In the following we assume that one of the order parameters stands for a melt phase.
The tilt function $g(\{\phi_i\})$, which we do not specify here yet, interpolates between the liquid and the solid;
particular choices will be discussed later.
Therefore, deviations of the temperature $T$ from the melting temperature $T_M$ favour energetically either the solid or liquid phase.
This term also contains the latent heat $L$ (dimension: energy/volume).

A specific feature of the model is that the evolution of the microstructure is expressed in terms of interface fields
\begin{equation} \label{intro::eq4}
\dot{\psi}_{\alpha\beta} := - \left( \frac{\delta}{\delta \phi_\alpha} - \frac{\delta}{\delta \phi_\beta} \right) F,
\end{equation}
such that the phase field evolution in the interface regions, $0< \phi_\alpha < 1$, reads
\begin{equation} \label{intro::eq5}
\dot{\phi}_\alpha = \frac{1}{\tilde{N}} \sum_{\beta\neq \alpha} \mu_{\alpha\beta} \dot{\psi}_{\alpha\beta},
\end{equation}
with kinetics coefficients $\mu_{\alpha\beta}=\mu_{\beta\alpha}>0$.
Here, $\tilde{N}$ is the number of phases with non-vanishing volume fractions at the present position.
Since this number is different in situations with a binary interface, a triple junction or overlapping interfaces, it is obvious that also the interface profiles are determined in a piecewise manner.
These regions have to be connected by appropriate boundary conditions, as discussed in detail in \cite{Bhogireddy:2014aa}.

\section{Grain boundary wetting}
\label{wettingsection}

\subsection{Analytical solution}
\label{multi:section}

\begin{figure}
\begin{center}
\includegraphics[width=8cm]{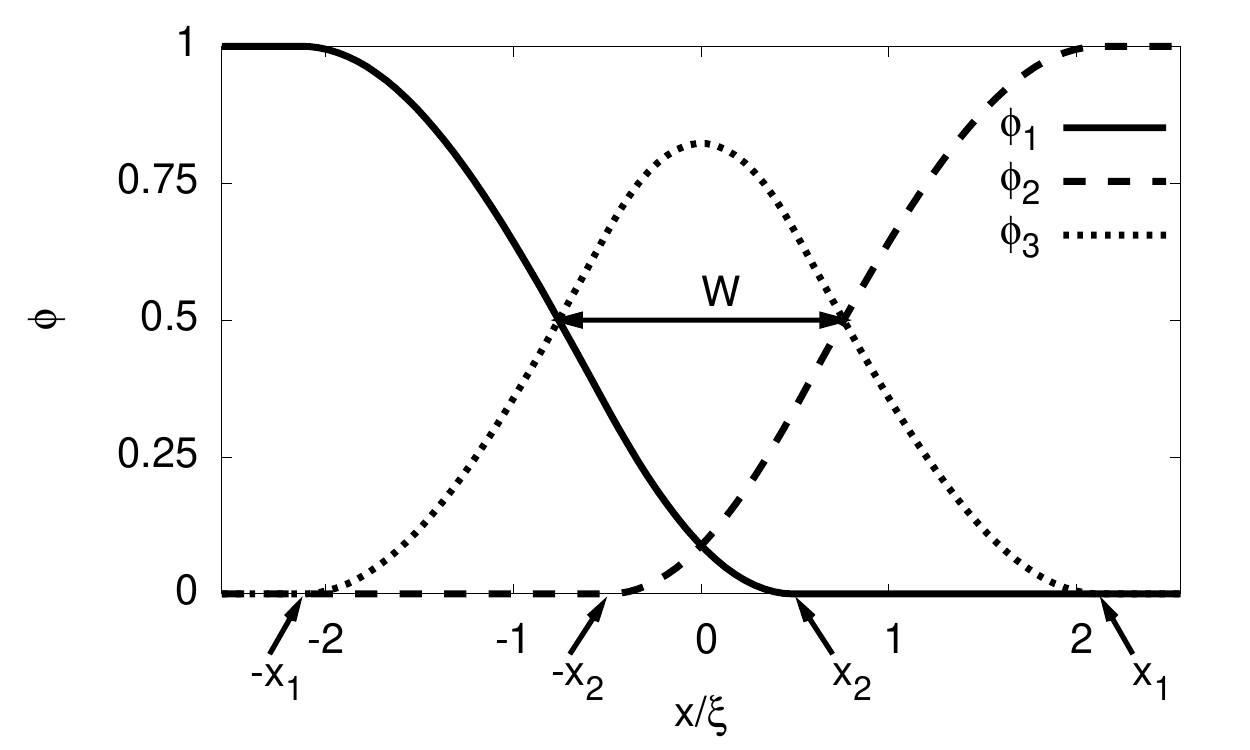}
\caption{Overlapping phase field profiles, as obtained from the analytical solution of the problem.
Parameters are $\sigma_{gb}/\sigma_{sl}=3$ and $\dimtemp_{13}=0.25$.
The melt layer thickness $W$ is defined as the distance between the points where $\phi_3=1/2$.
}
\label{fig6}
\end{center}
\end{figure}

As discussed in the introduction, the premelting at a grain boundary in a phase field model appears via the overlap of the interface profiles.
Such a situation is shown in Fig.~\ref{fig6}, where we use three order parameters to distinguish the phases and grains.

In particular, we use here $\phi_1$ and $\phi_2$ for the two distinct solid grains and $\phi_3$ for the melt phase.
Planar interfaces are assumed in Fig.~\ref{fig6}, therefore the phase fields have a nontrivial dependence only in the interface normal direction ($x$ direction) and are translational invariant in the others.
Effectively, the problem therefore becomes one-dimensional.
As discussed in \cite{Bhogireddy:2014aa} it is not sufficient to use a single order parameter phase field model, which only distinguishes between solid and melt, but not between the grains. 
In that case, there is no grain boundary energy associated with the interface between them, and the interaction of the two solid-melt interfaces is always attractive.
This implies that grain boundary premelting does not occur in such a situation.
In contrast, for the multi-order parameter case, the premelting transition is directly linked to the ratio of the grain boundary energy to the solid-melt interfacial energies and correctly captured by the phase field model \cite{Bhogireddy:2014aa}.

In the following we briefly summarise the steps to solve this problem analytically for a specific choice of the thermal coupling function $g(\{\phi_\alpha\})$, which serves as a benchmark for the comparison with phase field simulations using {\sc OpenPhase} in the following subsection.
There, also other choices of the coupling function will be discussed.
The particular case 
\begin{equation} \label{linearcouplingfunction}
g(\phi_1, \phi_2, \phi_3)=1-\phi_3
\end{equation}
has the advantage that the equilibrium phase field equations are linear, and therefore their solution is straightforward.

As shown in Fig.~\ref{fig6} the phase fields separate the entire domain into three different regions, and the phase field equations have to be solved piecewise in these domains.
For symmetrical grain boundaries with $\sigma_{sl}=\sigma_{13}=\sigma_{23}$ (still different from the grain boundary energy $\sigma_{gb}=\sigma_{12}$ \cite{Read:1950fk}) and equal interface thicknesses $\eta_{\alpha\beta}=\eta$ the situation is symmetrical around the center point $x=0$.
First, in the single phase domain $x<-x_1$ trivially $\phi_1=1$ and $\phi_2=\phi_3=0$.
Second, in the region $-x_1<x<-x_2$ the two phase fields for the left grain and the melt have nontrivial values, whereas still $\phi_2\equiv0$.
Hence $\phi_3=1-\phi_1$ there.
The equilibrium phase field equations therefore reduce to 
\begin{equation}
-2\xi^2\phi_1'' + 1-2\phi_1 + \dimtemp_{13} = 0,
\end{equation}
with $\dimtemp_{13}=L\eta(T-T_M)/[4\sigma_{sl}T_M]$ as a measure for the undercooling or overheating.
The interface thickness is given by $\xi=\eta/\pi$ and the dash is used to indicate the spatial derivative $\partial/\partial x$.
The proper solution of this equation is 
\begin{equation}
\phi_1(x)= \frac{1+\dimtemp_{13}}{2} + A\sin\frac{x+x_0}{\xi}.
\end{equation}
with integration constants $A$ and $x_0$.

Third, in the region $|x|<x_2$ all phase fields contribute, and it is the overlap of the two solid phase order parameters which induces the solid-melt interface interaction.
There the equilibrium phase field equations read
\begin{eqnarray}
&&1-2\phi_1- \left(2-\frac{\sigma_{gb}}{\sigma_{sl}} \right) \phi_2 - 2\xi^2\phi_1'' \nonumber \\
&&-\xi^2\left(2-\frac{\sigma_{gb}}{\sigma_{sl}}\right)\phi_2'' =- \dimtemp_{13} 
\end{eqnarray}
and
\begin{eqnarray}
&&1-2\phi_2 - \left(2-\frac{\sigma_{gb}}{\sigma_{sl}} \right) \phi_1 - 2\xi^2\phi_2'' \nonumber \\
&&-\xi^2\left(2-\frac{\sigma_{gb}}{\sigma_{sl}}\right)\phi_1'' = -\dimtemp_{13},
\end{eqnarray}
with the solutions
\begin{eqnarray}
\phi_1(x) &=& \frac{1+\dimtemp_{13}}{4-\frac{\sigma_{gb}}{\sigma_{sl}}} + B\sin\frac{x+y_0}{\xi}, \label{lab1} \\
\phi_2(x) &=& \frac{1+\dimtemp_{13}}{4-\frac{\sigma_{gb}}{\sigma_{sl}}} - B\sin\frac{x-y_0}{\xi}. \label{lab2} 
\end{eqnarray}
Here we have already considered the symmetry $\phi_1(x)=\phi_2(-x)$.
Again, $y_0$ and $B$ are integration constants.

\begin{figure}
\begin{center}
\includegraphics[width=8cm]{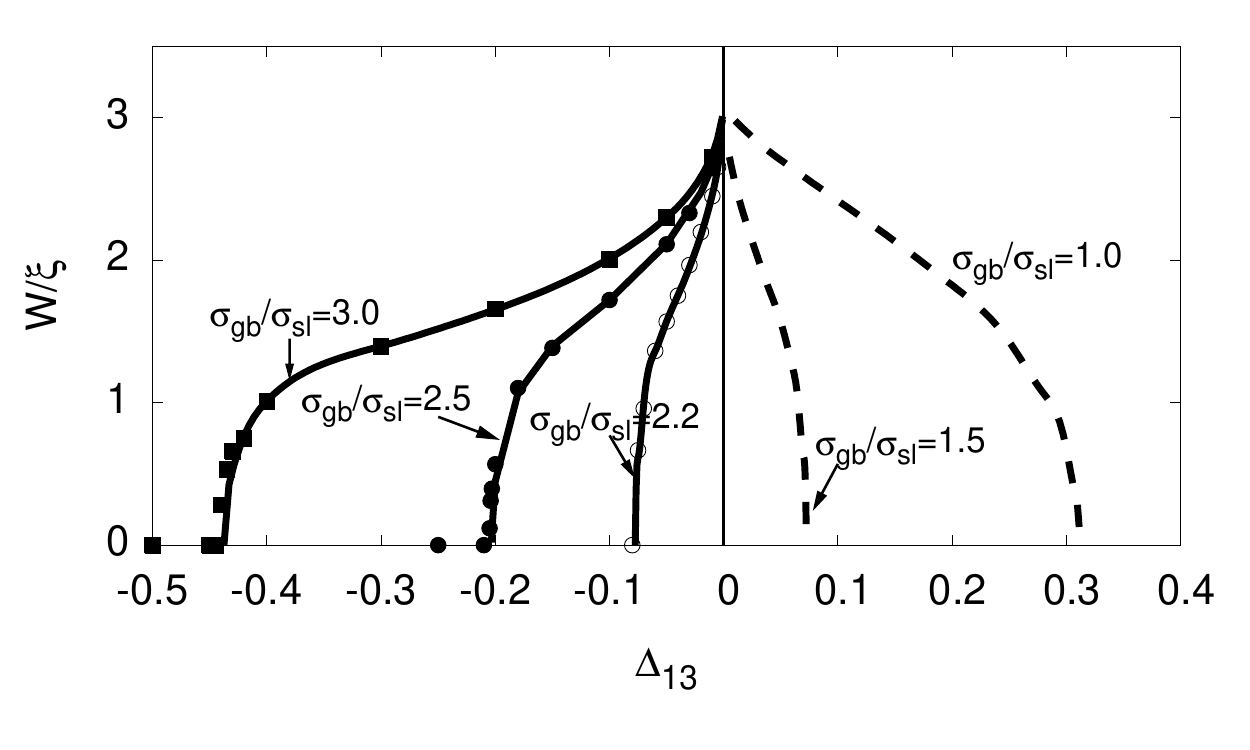}
\caption{Melt layer thickness as function of temperature for different ratios of the interfacial energies. 
The continuous curves result from the analytical solution for different values of $\sigma_{gb}/\sigma_{sl}$.
For the repulsive cases $\sigma_{gb}/\sigma_{sl}>2$ also results from numerical {\sc OpenPhase} simulations are shown.
They agree perfectly with the analytical solution.
The transition between attractive and repulsive situations occurs for $\sigma_{gb}=2\sigma_{sl}$.
}
\label{fig8}
\end{center}
\end{figure}

As elaborated in \cite{Bhogireddy:2014aa}, the continuity of the phase fields and their derivatives at the connecting points $x_1$ and $x_2$ is nontrivial and follows from energy minimisation.
In the end one arrives at the conditions
\begin{eqnarray}
x_0 &=&  x_1 + \frac{\pi \xi}{2}, \quad
A = \frac{1-\dimtemp_{13}}{2}, \\
y_0 &=& -x_2 - \frac{\pi\xi}{2}, \quad
B = \frac{1+\Delta_{13}}{4-\sigma_{gb}/\sigma_{sl}}.
\end{eqnarray}
The location of the connection points $x_1$ and $x_2$ are determined by the relations
\begin{equation}
A\sin\frac{x_1-x_2}{\xi} = B \sin\frac{2x_2}{\xi}
\end{equation}
and
\begin{eqnarray}
&& \frac{1+\dimtemp_{13}}{2} + A\cos\frac{x_1-x_2}{\xi} \nonumber \\
&=& \frac{1+\Delta_{13}}{4-\sigma_{gb}/\sigma_{sl}}- B \cos\frac{2x_2}{\xi}.
\end{eqnarray}
From this solution we can compute the melt layer thickness $W$ as function of temperature.
Here we define $W$ as the distance between the points, where the melt order parameter $\phi_3$ has the value $\phi_3=1/2$, see Fig.~\ref{fig6}.
The result is shown in Fig.~\ref{fig8} for different ratios of the grain boundary to the solid-melt interfacial energy.

\begin{figure}
\begin{center}
\includegraphics[width=8cm]{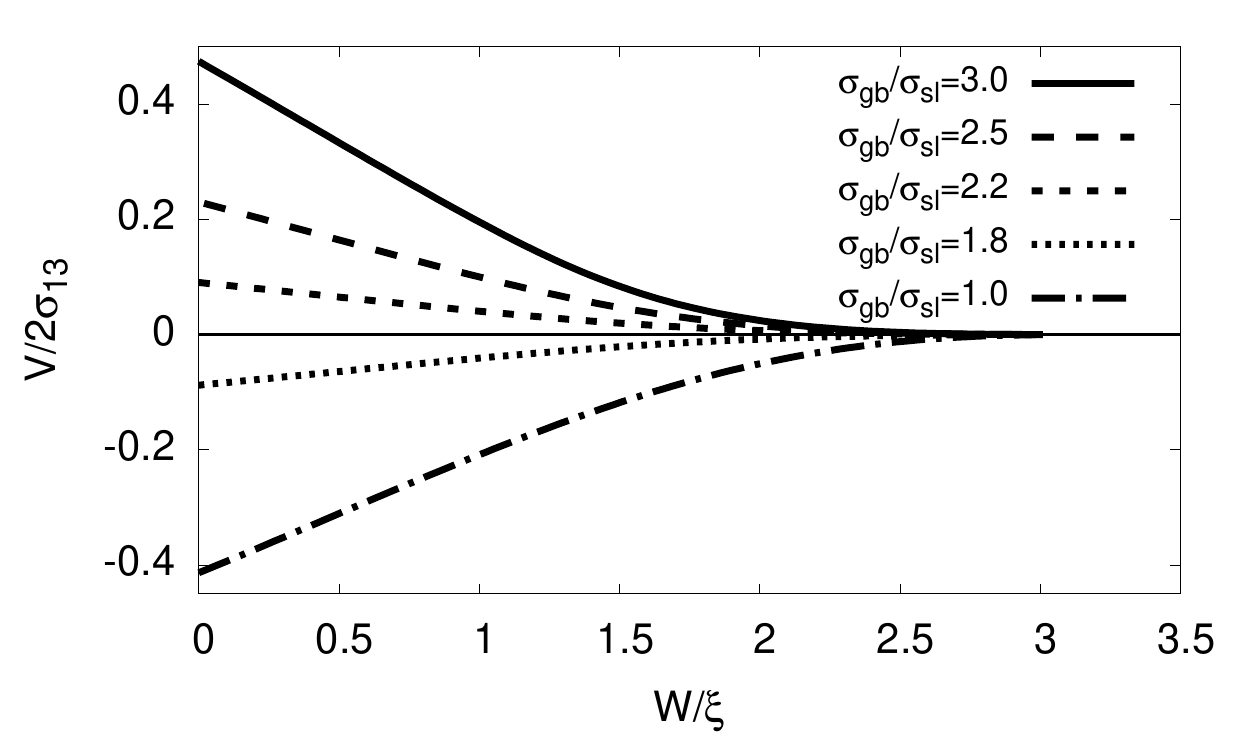}
\caption{Analytically calculated disjoining potential as function of the melt layer thickness for the multi-order parameter phase field model.
}
\label{fig9}
\end{center}
\end{figure}

From the calculated free energy per unit area of the system $\cal F$  the solid-melt interface interaction (disjoining potential $V(W)$) can be determined according to the relation
\begin{equation} \label{eq2}
\frac{V(W)}{2\sigma_{sl}} = \frac{{\cal F}}{2\sigma_{sl}} + \frac{2}{\pi} \dimtemp_{13} \frac{W}{\xi} - 1.
\end{equation}
The result is shown in Fig.~\ref{fig9}.
The transition from attractive to repulsive interactions takes place at $\sigma_{gb}=2\sigma_{sl}$.
The interaction becomes strictly zero for $W/\xi>\pi$, as then the two order parameters for the solid phases no longer overlap.

\subsection{{\sc OpenPhase} modeling}
\label{OPFGBpremelting}


To complement the previous analytical predictions we use numerical simulations to confirm them.
For that, we use the open source phase field package {\sc OpenPhase} \cite{openphase}, which is based on the phase field model introduced in section \ref{model::section}.
As of today, this package allows to simulate various problems in materials science, including solidification, solid state transformations, grain growth etc., taking into account diffusive and hydrodynamic transport and advection, elastic and plastic deformations.
The different physical ingredients are encapsulated in separate modules, which can be used according to the needs of the problem of interest.
One of the major strengths of the package is that it can deal with an arbitrary number of phases or grain orientations, which are encoded by separate phase fields.
Since in most regions of the computational domain only one phase field has a nontrivial value, apart from dual interfaces and triple junctions, where two or three phases come together, a smart memory management is used that only the non-vanishing phase field values are stored.
This allows to perform large scale simulations at a low memory consumption, which would otherwise strongly limit the maximum system size.
The code is written in C++ in an object oriented way and is fully parallelized using {\sc OpenMP} \cite{openmp}.

For the present problem of grain boundary wetting with planar interfaces the problem becomes one-dimensional.
In interface normal direction the system size has to be at least several times the width of the phase field interface thickness.
In contrast to simulations of e.g.~solidification, where the interface thickness serves only as numerical parameter and should be as small as possible in comparison to the physical length scales, here the finite thickness of the interface is central.
As we have shown before, it is the overlap of nearby surfaces which induces an interaction between them.
Hence the interface thickness is here assumed to be of the order of a nanometer, in contrast to the aforementioned microstructure evolution problems, where it is often chosen to be of the order of a micrometer and therefore significantly thicker than the true physical interface thickness.
It is important to properly discretize the interface, and we typically use $\eta/\Delta x=20$, with $\Delta x$ being the grid spacing.

The simulations are run until convergence to equilibrium is reached.
Since we use here the dynamics given by Eq.~(\ref{intro::eq5}), only stable equilibrium solutions can be found this way.
Cases with attractive interfaces, however, correspond to unstable solutions for $T>T_M$, and therefore we focus here on repulsive interactions for $\sigma_{gb}/\sigma_{sl}>2$.

From the resulting phase field interface profiles the melt layer thickness $W$ is extracted.
As a first benchmark we aim here at a comparison with the above analytical investigation, which are based on the coupling function (\ref{linearcouplingfunction}).
Since this is not the default choice in the {\sc OpenPhase} code, appropriate adjustments have to be made.
The results are shown in Fig.~\ref{fig8} together with the analytical predictions, and we find an excellent agreement between them.
Below a certain ``bridging temperature'', which depends on the interfacial energy ratio $\sigma_{gb}/\sigma_{sl}$, the melt layer thickness becomes zero and the grain boundary is dry.
Here, however, one has to keep in mind that with the present choice of the definition of $W$ this means that $\phi_3<1/2$, but does not necessarily imply that the liquid phase field is vanishing everywhere.
The choice of a different measure of the melt layer thickness, which is based on an integral expression across the interface using $\phi_3$, is sensitive also to values $\phi_3<1/2$.
This definition and its consequences are discussed in detail in \cite{Bhogireddy:2014aa}.
In particular, the bridging temperature, below which the grain boundary is strictly dry, can easily be predicted analytically in this case.

Apart from the thermal coupling function (\ref{linearcouplingfunction}), which is convenient for analytical considerations, we can also use different choices.
In particular, {\sc OpenPhase} natively uses a different one, which is given by
\begin{eqnarray}
g(\phi_3) &=&  -\frac{1}{4} \Big[(2\phi_3-1)\sqrt{\phi_3(1-\phi_3)} \nonumber \\
&&+ \frac{1}{2} \arcsin(2\phi_3-1)\Big]. \label{Openphasecoupling}
\end{eqnarray}
This function has the advantage that it maintains the same interface profile for a moving as for a stationary interface \cite{steinbach}.
It interpolates monotonically between $g(0)=\pi/16\approx 0.196$ and $g(1)=-\pi/16\approx -0.196$, which requires to renormalise the driving force, which multiplies this function,
\begin{equation}
L \frac{T-T_M}{T_M} \to \frac{8}{\pi} L \frac{T-T_M}{T_M},
\end{equation}
in order to have the same temperature definition as before.
In the equations of motion the derivative of the coupling function,
\begin{equation}
\frac{\partial g}{\partial x_3} = -\sqrt{\phi_3 (1-\phi_3)}
\end{equation}
appears.

A second difference is the use of the driving force, $\Delta f = L(T-T_M)/T_M$, as implemented in {\sc OpenPhase}.
To avoid numerical instabilities in the code for too large driving forces, it is regularised by the replacement
\begin{equation} \label{cutoff}
\Delta f \to \Delta f_{max} \tanh\frac{\Delta f}{\Delta f_{max}}
\end{equation}
with a cutoff value $\Delta f_{max}=2\pi \chi \sigma_{sl}/\eta \tilde{N}$, with an adjustable dimensionless parameter $\chi$.
Obviously, for $|\Delta f|\ll \Delta f_{max}$ both expressions coincide.
In contrast, for $|\Delta f|\gg \Delta f_{max}$ the driving force saturates, which in our case can prevent interfaces from becoming completely dry in the low temperature regime.

\begin{figure}
\begin{center}
\includegraphics[width=8cm]{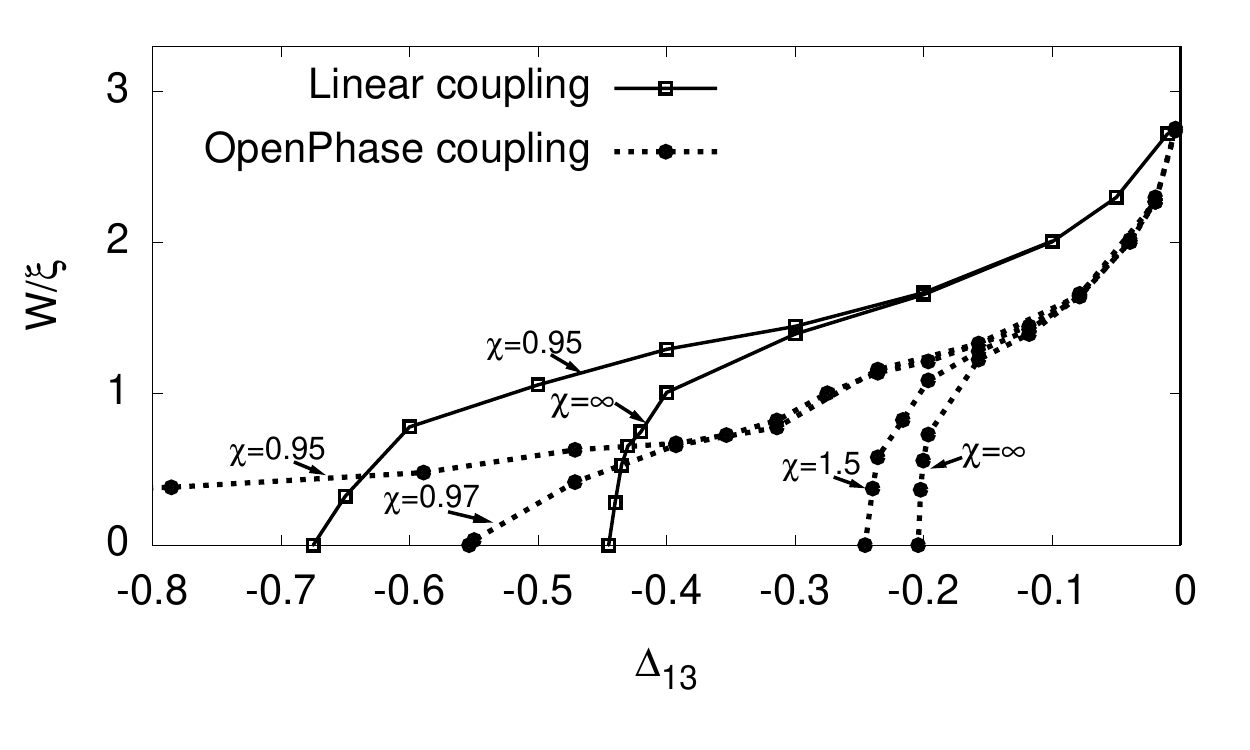}
\caption{
Melt layer thickness as function of the dimensionless temperature deviation $\Delta_{13}$ from the melting temperature, for $\sigma_{gb}/\sigma_{sl}=3$.
Different thermal coupling functions are used here.
For linear coupling according to Eq.~(\ref{linearcouplingfunction}) the analytical and numerical results agree.
Additionally, the {\sc OpenPhase} coupling function (\ref{Openphasecoupling}) is used.
Both functions are used without and with cutoff according to expression (\ref{cutoff}) with different cutoff parameters $\chi$.
}
\label{fig4}
\end{center}
\end{figure}

For comparison with the above results, we therefore additionally run simulations which use the linear coupling function (\ref{linearcouplingfunction}), but with the cutoff (\ref{cutoff}).
The results are shown in Fig.~\ref{fig4}.
As expected, for low driving forces, the melt layer thickness agrees with the previous results.
For stronger undercoolings, it is cut off, and therefore the melt layer thickness persists to lower temperatures.
The cutoff can therefore be used to adjust the bridging temperature.
Furthermore, we also used the coupling function (\ref{Openphasecoupling}), with and without the cutoff (\ref{cutoff}).
The results are also shown in Fig.~\ref{fig4}.
By varying the cutoff parameter $\chi$ we can shift the bridging temperature at which the melt layer thickness becomes zero, and even can let a melt layer remain at very low temperatures.
Hence, tuning the coupling function $g$ and the cutoff $\chi$ can be used to adjust the phenomenological phase field model to specific materials and grain misorientations.

\section{Melting along grain boundaries}
\label{kinetics::section}

\subsection{Phase field modelling}
\label{PFmelting::subsection}

\begin{figure}
\begin{center}
\includegraphics[trim=8.4cm 9.3cm 8.4cm 9.3cm, clip=true, width=7cm]{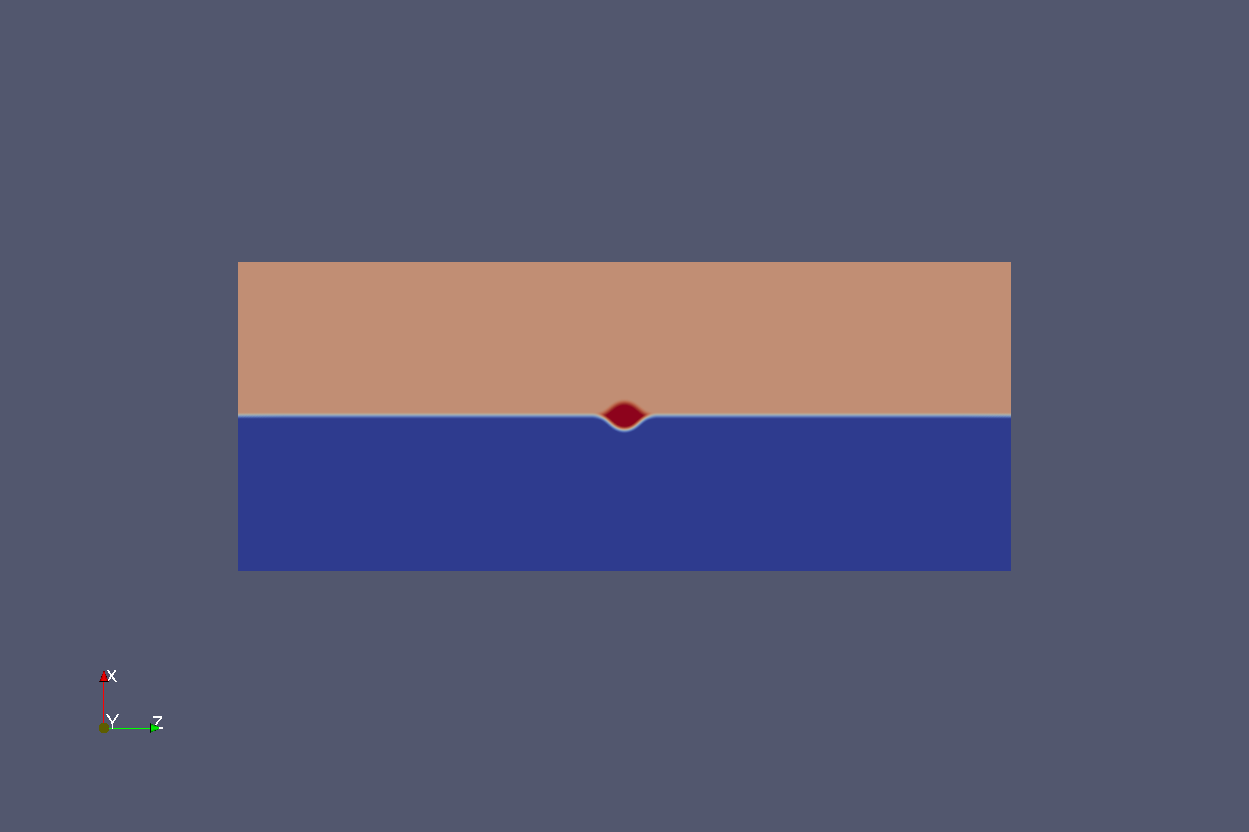}\\[1ex]
\includegraphics[trim=8.4cm 9.3cm 8.4cm 9.3cm, clip=true, width=7cm]{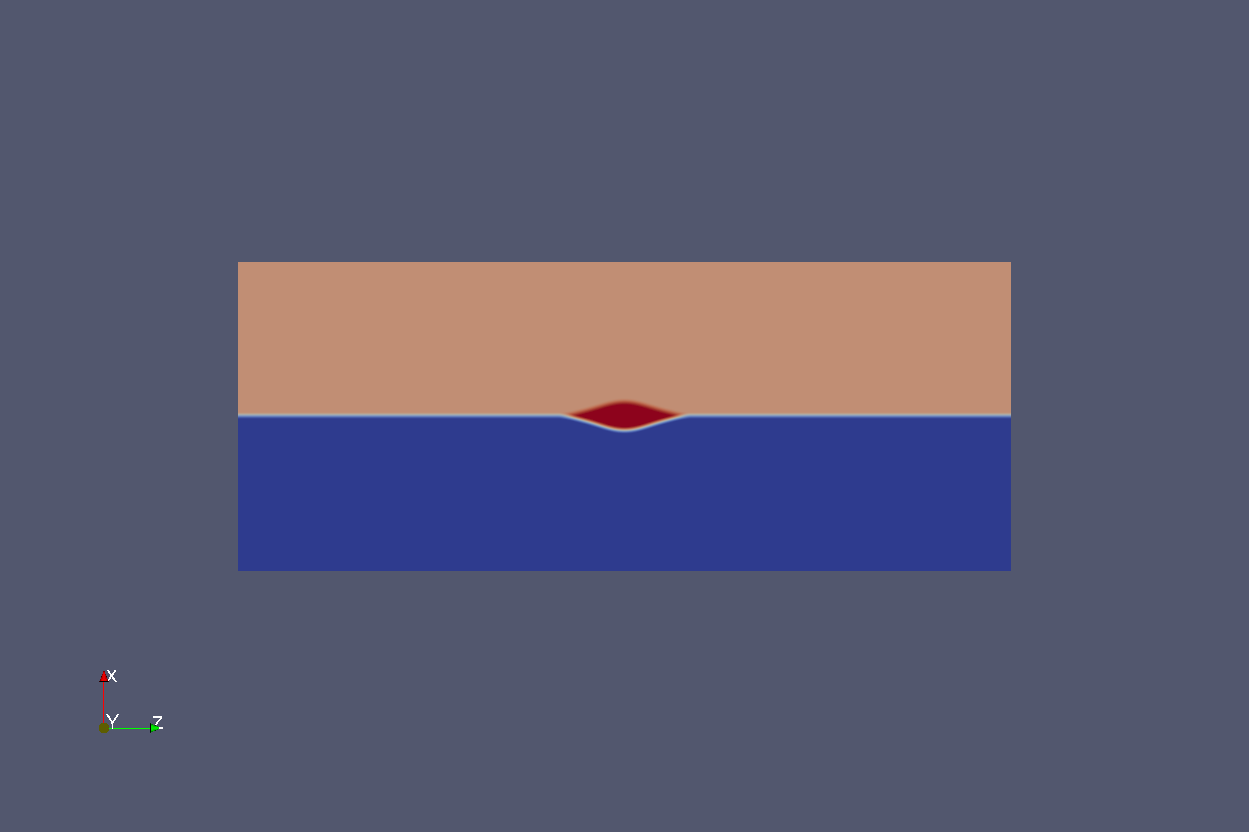}\\[1ex]
\includegraphics[trim=8.4cm 9.3cm 8.4cm 9.3cm, clip=true, width=7cm]{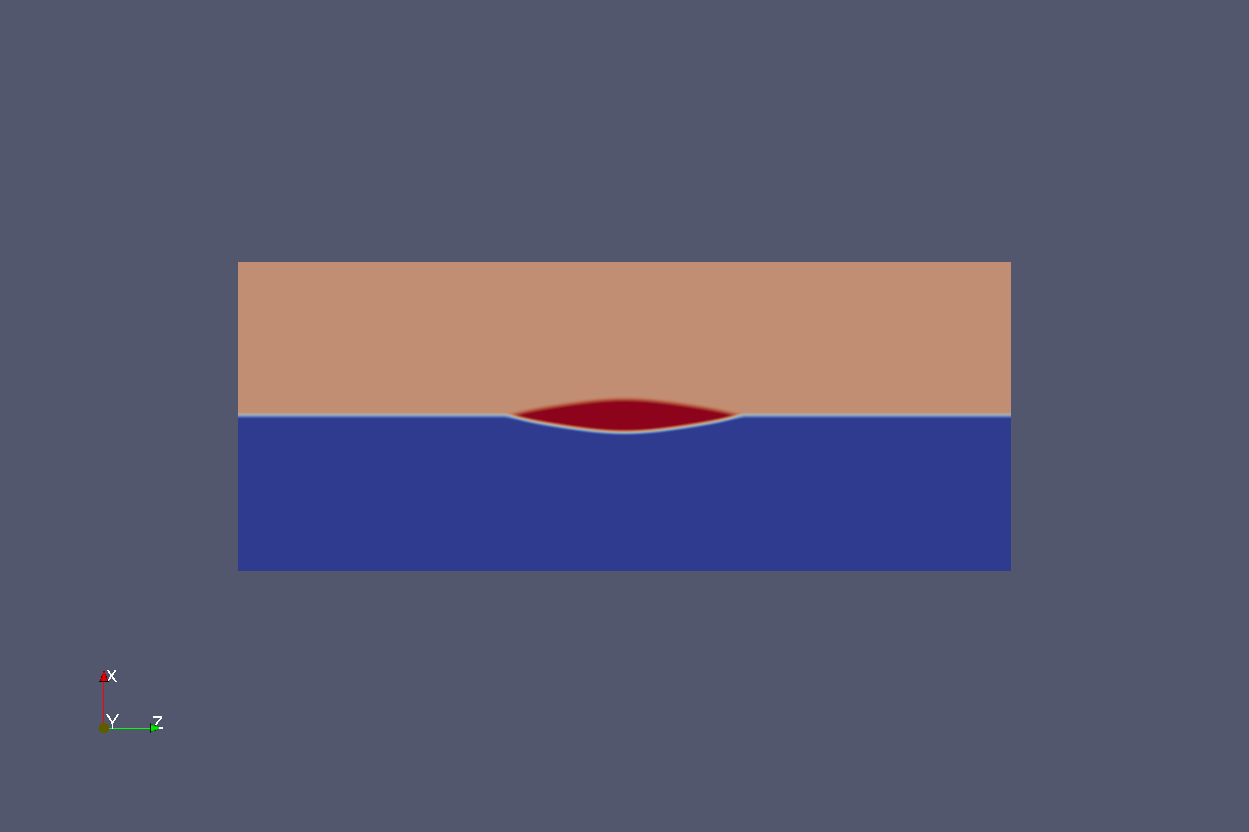}\\[1ex]
\includegraphics[trim=8.4cm 9.3cm 8.4cm 9.3cm, clip=true, width=7cm]{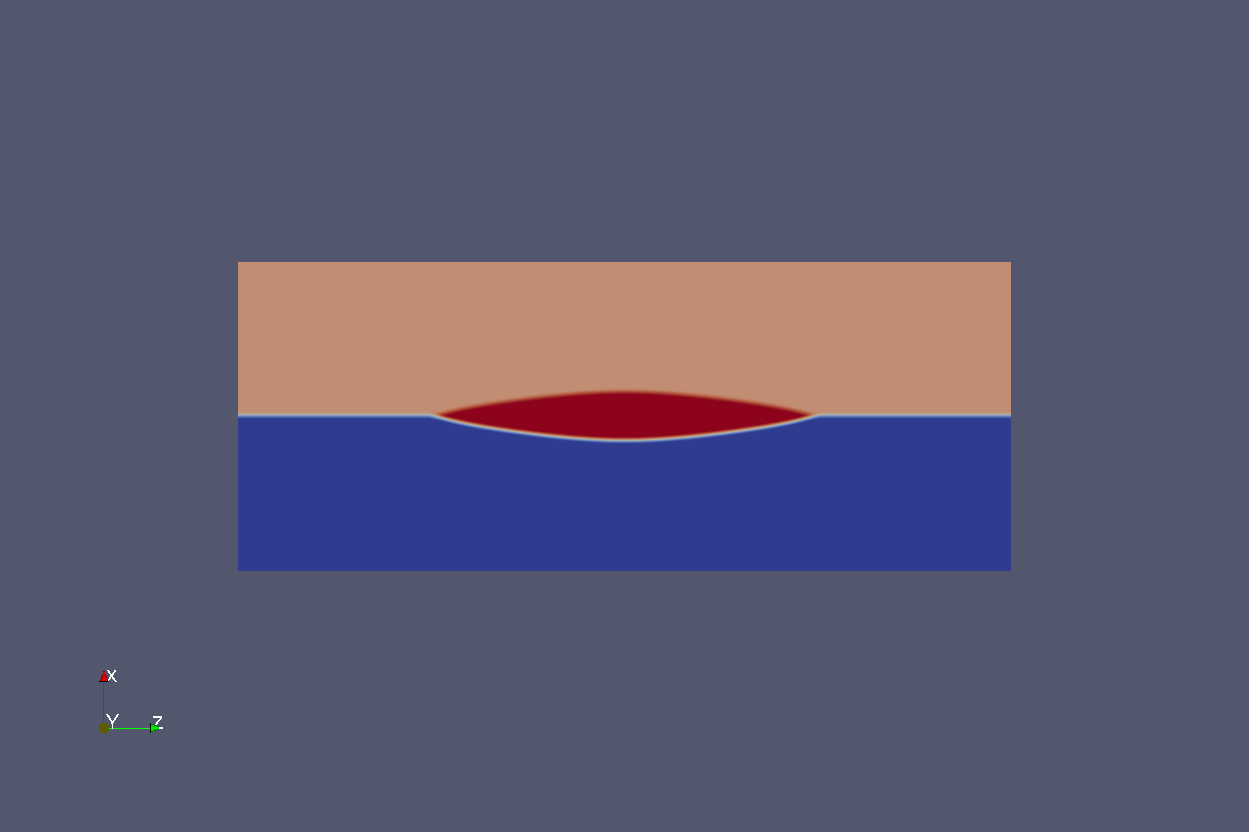}
\caption{Snapshots of two-dimensional phase field simulations of the melting process along a horizontal dry grain boundary.
The different colours discriminate between the grains and phases.
Initially, a spherical nucleus is placed on the grain boundary, which then spreads laterally.
The parameters are in dimensionless form (see section \ref{sharpinterfacemelting::subsection} for definitions) $\sigma_{gb} /\sigma_{sl} =1.9$, $\Delta=0.3$, $\eta/d = 0.8$, $L/c_pT_M = 0.2$, $\mu_{\alpha\beta}\sigma_{sl}/D= 1.54$. 
The system size is $80.08\, d\times 32.08\, d$, and at the boundaries the constant overheating is applied.
The lattice spacing is $\Delta x/d = 0.08$, the time step is $D\Delta t/d^2=8.3\cdot10^{-4}$. 
The snapshots are taken at $t D/d^2=0$, $t D/d^2=7.9$, $t D/d^2=82.78$ and $t D/d^2=332.4$ (from top to bottom).}
\label{figPF1}
\end{center}
\end{figure}

\begin{figure}
\begin{center}
\includegraphics[width=8cm]{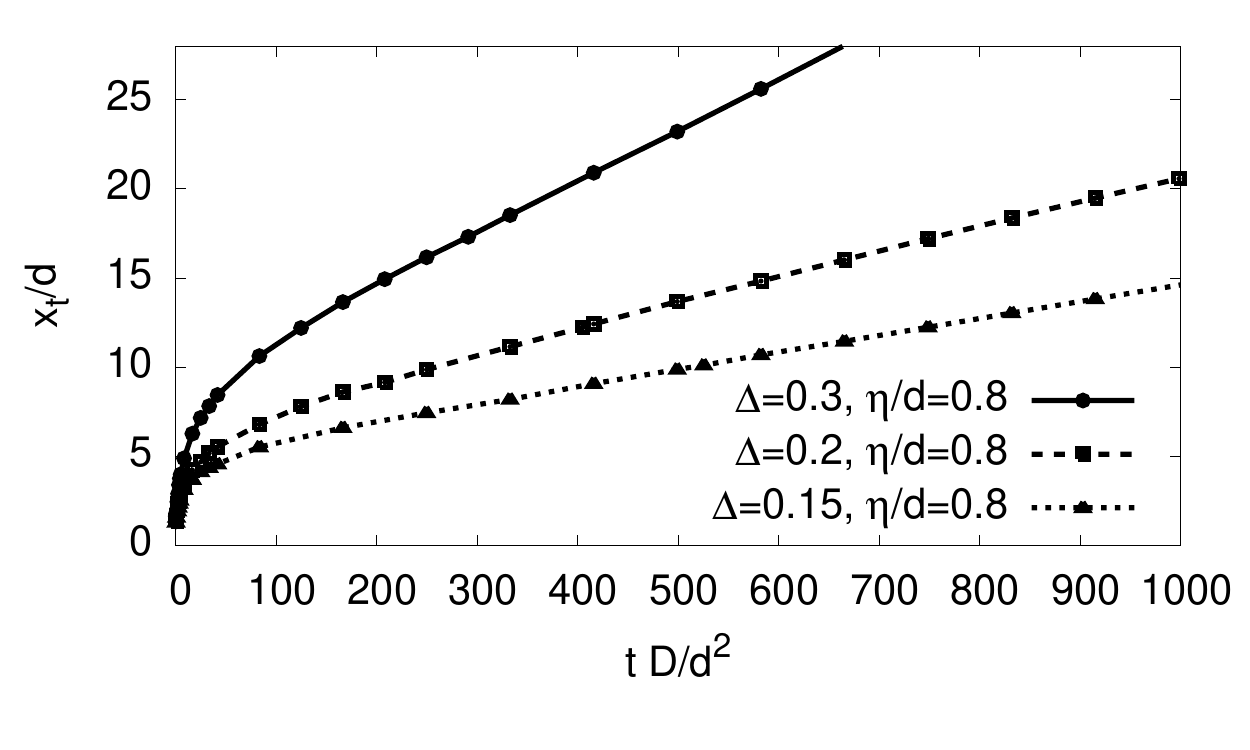}
\caption{Lateral trijunction position as function of time, expressed through the capillary length $d=T_M\sigma_{sl}c_p/L^2$.
The results stem from the phase field simulations.
The parameters are the same as in Figure \ref{figPF1} apart from those given explicitly in the legend.
}
\label{figPF2}
\end{center}
\end{figure}

To understand the grain boundary melting kinetics, we start our investigations with phase field simulations.
For that, we set up a two-dimensional situation as depicted in Fig.~\ref{figPF1}.
A liquid nucleus is placed on a dry grain boundary, and the system is subjected to an overheating $T_\infty>T_M$ from far away, which is also the homogeneous initial temperature in the system.
This triggers a melting process which consumes latent heat at the moving interfaces and therefore leads to an inhomogeneous temperature distribution in the system.
We therefore have to supplement the previous phase field equations by the diffusion equation for the temperature field,
\begin{equation}
\frac{\partial T}{\partial t} = D\nabla^2 T - \frac{L}{c_p} \frac{\partial\phi_3}{\partial t}.
\end{equation}
We use here the coupling function (\ref{Openphasecoupling}) without the cutoff (\ref{cutoff}).
We concentrate here on a symmetrical model, where the heat diffusivity $D$ and the heat capacity $c_p$ are the same in both phases.
Furthermore, we focus here on attractive interactions between the solid-melt interfaces, hence $\sigma_{gb} < 2\sigma_{sl}$.
In the course of time the nucleus spreads along the grain boundary and establishes a mesoscopic contact angle $\phi_\infty$ at the triple junction (see also Fig.~\ref{geometry} for the definition of this angle).
This contact angle is defined on a scale much larger than the interface thickness and must not be confused with a microscopic contact angle $\phi_0$, which is defined in the region of the overlapping solid-melt interfaces, and which will be discussed in more detail in the following subsection.
In equilibrium, we expect Young's law to hold, thus
\begin{equation}
\sigma_{gb} = 2\sigma_{sl} \cos \phi_\infty^{eq}.
\end{equation}
For dynamical situations we find slight deviations from this relation, which also affect the microscopic contact angle $\phi_0$;
for a more general discussion of this issue we refer to \cite{Guo:2011aa}.
With the dimensionless overheating for this diffusion limited process being defined as $\Delta = (T_\infty-T_M)c_p/L$ we have extracted the trijunction position $x_t$ as function of time for different driving forces $\Delta$.
The results are shown in Fig.~\ref{figPF2}.
As expected, a higher overheating leads to faster melting.
Initially, the system shows a diffusive slowing down, which then crosses over to steady state growth, $x_t\sim t$.
We refrain here from a quantitative analysis of the latter regime using phase field methods and will discuss it in more detail in the following section, where sharp interface methods are applied.

\begin{figure}
\begin{center}
\includegraphics[width=8cm]{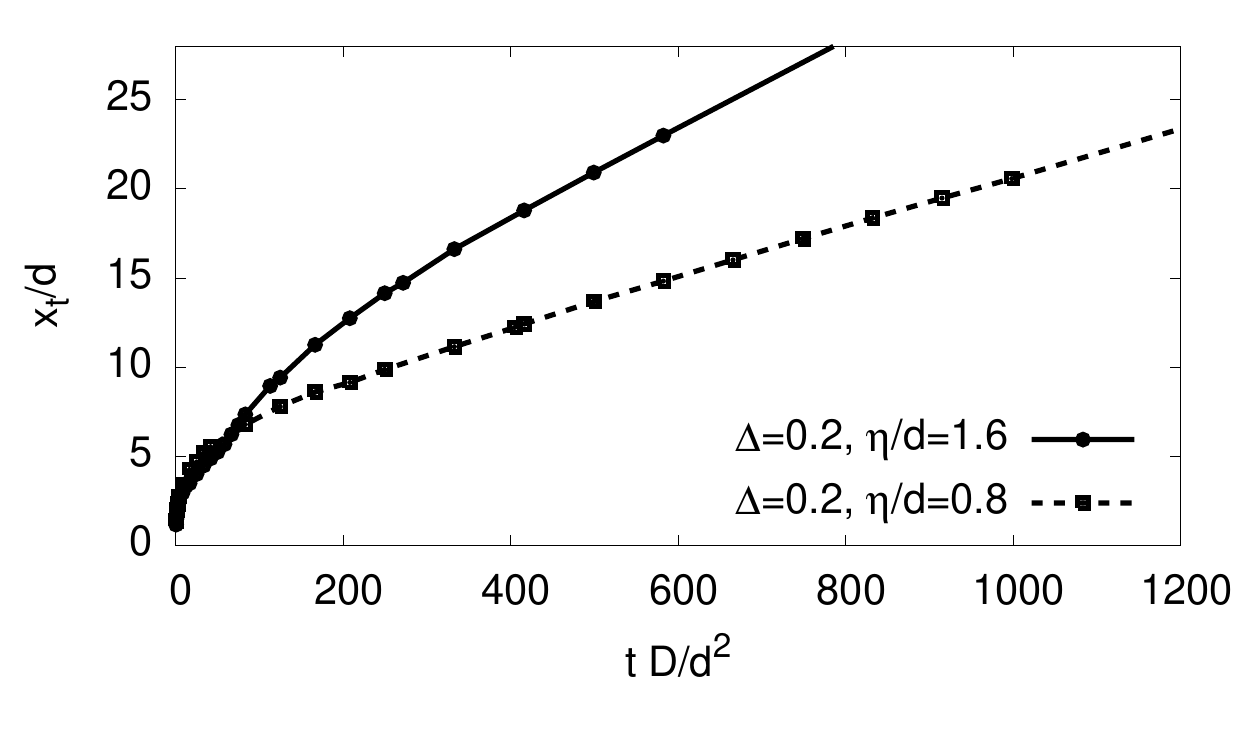}
\caption{Lateral trijunction position as function of time, expressed through the capillary length $d=T_M\sigma_{sl}c_p/L^2$.
The results stem from the phase field simulations.
The parameters are the same as in Figure \ref{figPF1} apart from those given explicitly in the legend. 
}
\label{figPF3}
\end{center}
\end{figure}

Another variation is to change the phase field interface thickness $\xi$ for fixed driving force $\Delta$.
Here we are particularly interested in the role of the short range interactions between nearby solid-melt interfaces.
Wider interfaces correspond to a longer-ranged interaction, as has been worked out in the previous section.
Especially for a small opening angle $\phi_\infty$ the overlap of the interfaces extends over a larger distance, and we can expect this to influence the melting kinetics. 
Indeed, we find that wider interfaces lead to faster growth, as shown in Fig.~\ref{figPF3}.
Since it is computationally demanding to explore this effect quantitatively in the framework of the present phase field model, as it is difficult to separate the role of the short ranged interactions from the deviation from the sharp interface limit, we leave such investigations for future simulations.
Instead, we use in the following subsection a sharp interface model to obtain quantitative predictions, where we can separate the effects.

\subsection{Sharp interface description}
\label{sharpinterfacemelting::subsection}

Complementary to the phase field modelling above we use here a sharp interface description of the diffusion limited melting along a dry grain boundary.
We concentrate here on the steady state regime, which is best accessible by Green's function methods.
The advantage in comparison to a phase field description is that we can rigorously incorporate local equilibrium at the moving solid-melt interfaces.
Therefore the effect of the finite interface thickness, which leads to deviation from this relation, can be separated from the influence of the short ranged solid-melt interface interactions.
In turn, these interactions need to be incorporated explicitly, as they no longer result from an overlap of the now infinitely thin interfaces.

We use here a disjoining potential $V(W)$, which is close to the one obtained from the multi-order parameter phase field model above.
To allow for analytical progress, instead of the numerically determined potential, we use an exponentially decaying interaction.
In agreement with the results above the grain boundary wetting transition is related to the parameter $\bar{\sigma} = \sigma_{gb} - 2\sigma_{sl}$ on the mesoscopic scale.
Hence, $\bar{\sigma} > 0$ belongs to the regime of repulsive interfaces and grain boundary premelting, whereas $\bar{\sigma} < 0$ corresponds to attractive grain boundaries. 
We write $V(W)= \bar{\sigma} f(W/\delta)$, where $\delta$ is the atomistic length scale characterising the range of the structural forces, and $f(W/\delta) = \exp (-W/\delta)$. 
In this sense, $\delta$ is directly related to the interface thickness $\xi$ in the phase field description.

\begin{figure}
\hspace{0.35cm}\includegraphics[width=7.65cm, height=4.5cm]{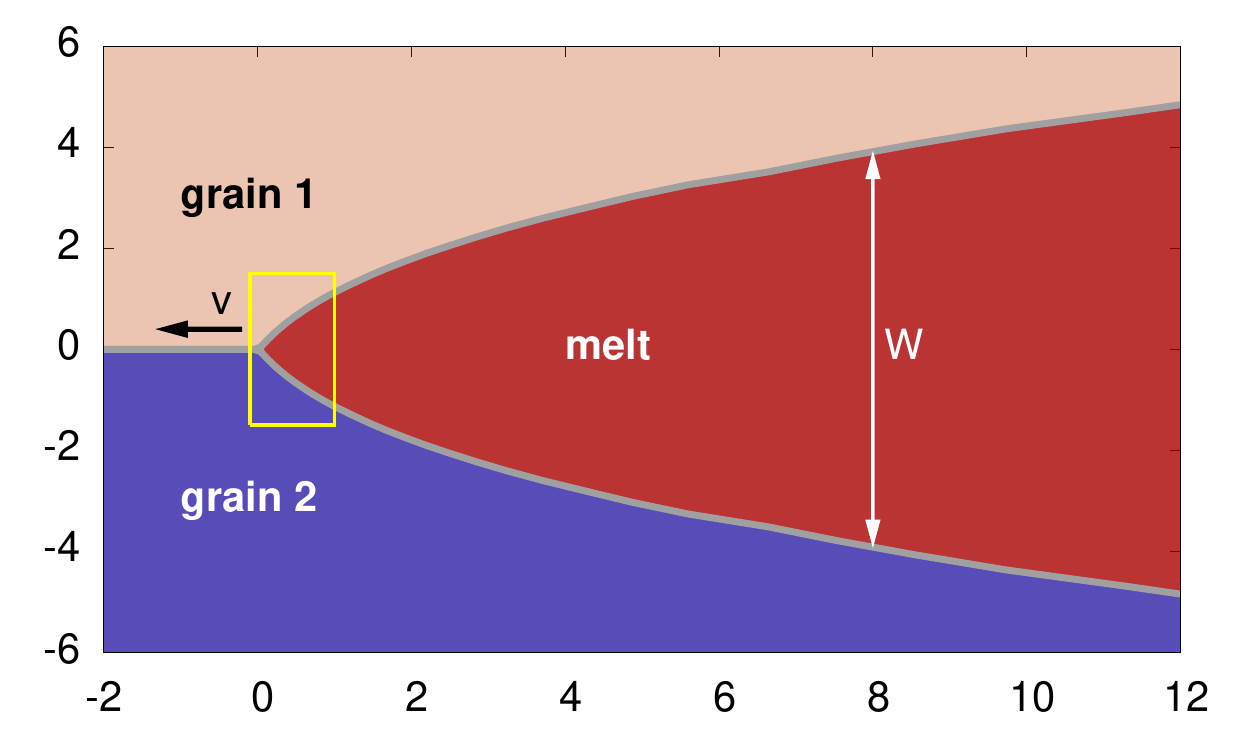}
\includegraphics[width=8cm, height=4.5cm]{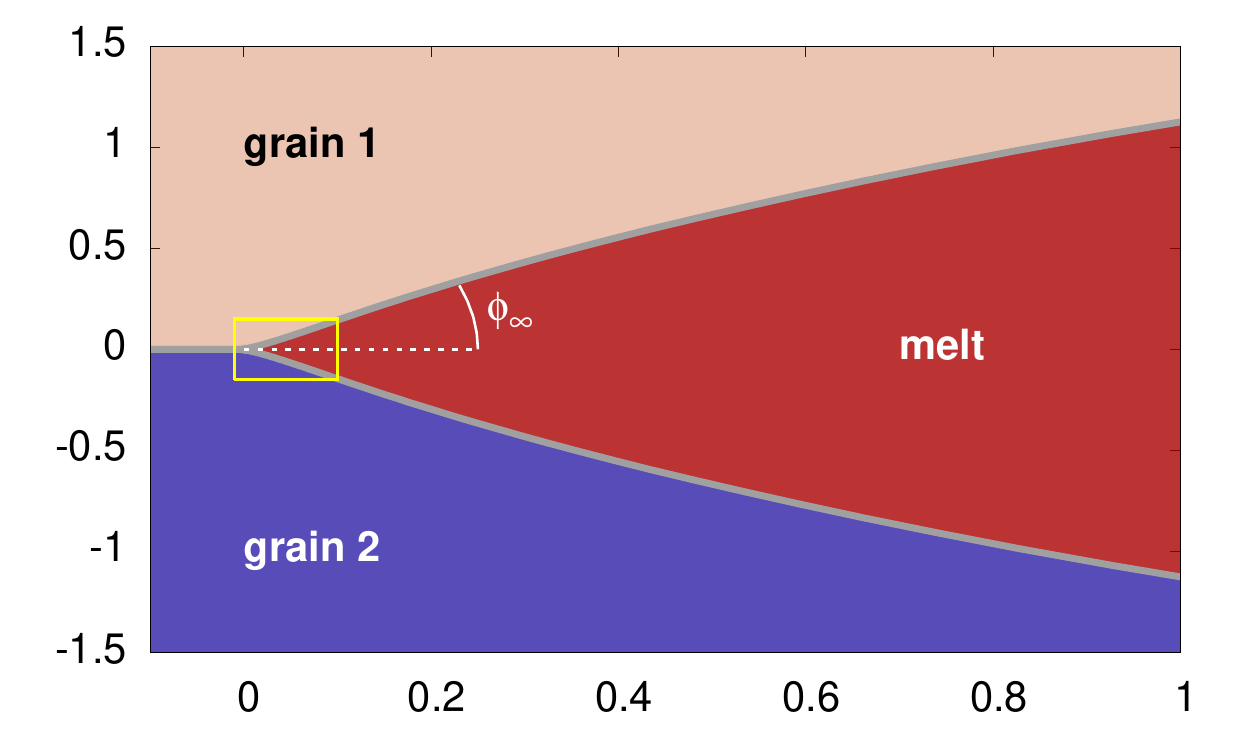}
\hspace*{-0.2cm}\includegraphics[width=8.2cm, height=4.5cm]{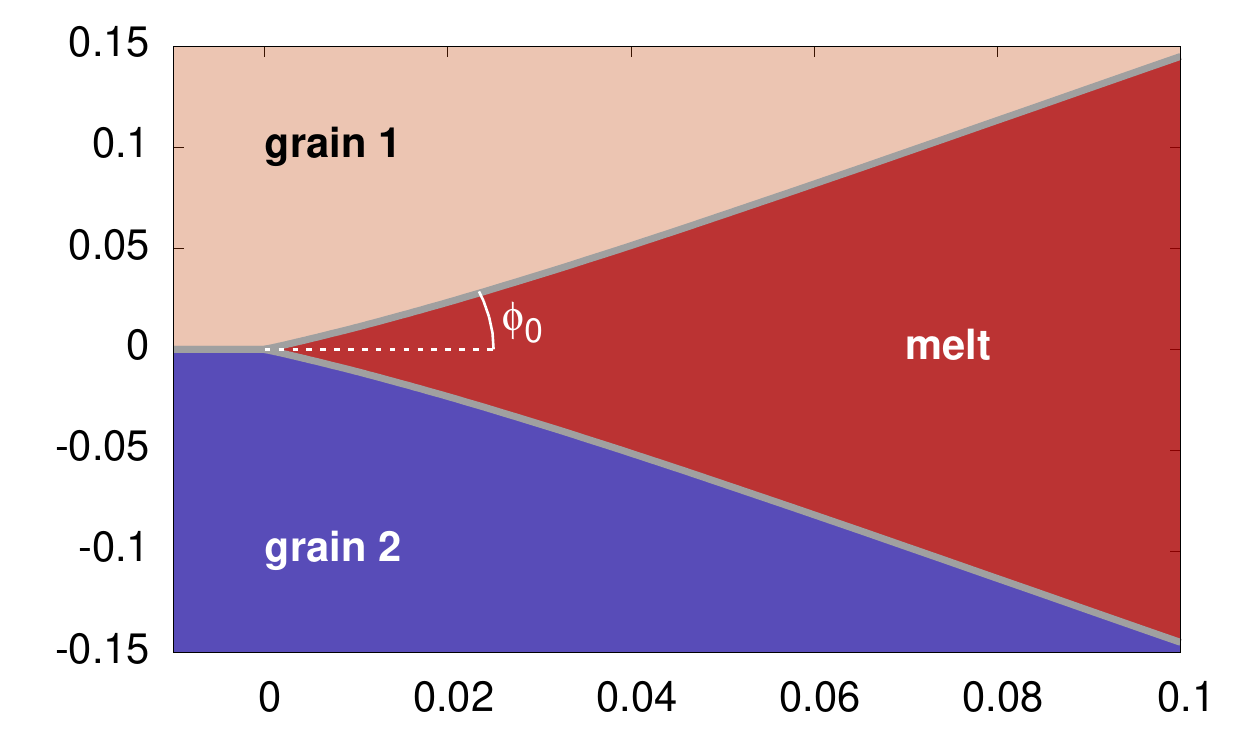}
\caption{\label{geometry}
Melting along an overheated (dry) grain boundary, as obtained from the sharp interface approach.
All lengths are measured in multiples of $\rho/\phi_0^2$ in the horizontal and $\rho/\phi_0$ in the vertical direction.
This unequal scaling implies that the opening angle appears larger than it is.
On the largest scale (top), the behavior is dominated by the diffusion limited growth, therefore asymptotically the melt contour approaches an Ivantsov parabola.
The regions inside the rectangles are magnified in the following subfigures. 
On intermediate scales the (mesoscopic) finite contact angle at the triple junction becomes relevant (centre), and differs from the angle $\phi_0$ on microscopic scales (bottom).
There the solid-melt interface interaction bends the interfaces.
Parameters are $\alpha=-5$, $\beta=4$, $\Delta=0.01$, $\phi_0=0.02$.
}
\end{figure}

The geometry of the system at the triple point is shown in Fig.~\ref{geometry}.
On large scales (top panel), a mesoscopic melt front is advancing along the dry grain boundary, with a parabolic front profile far behind the triple junction.
In the tip region (middle panel) a finite contact angles develops between apparently straight interfaces, in agreement with \cite{EAB_CH_DP_DET_PRL99_105701_2007}.
On the smallest scale (bottom panel) short ranged effects result in curved interface contours.

The short ranged interactions appear together with the Gibbs-Thomson correction for curved interfaces in the local equilibrium condition at the advancing solid-melt interfaces.
There, the equilibrium temperature is given by
\begin{eqnarray} \label{physicalLocEq}
T_{I} = T_M\Bigg[1 + \frac{\sigma_{sl} \kappa}{L} + \frac{\bar{\sigma}}{L \delta} f'(W/\delta) \Bigg],
\end{eqnarray}
with $\kappa$ being the curvature (positive for a convex liquid phase). 

For the further theoretical investigations it is advantageous to use a dimensionless representation.
Eq.~(\ref{physicalLocEq}) then reads
\begin{eqnarray} \label{dimLessLocEq}
u|_{int} = \Delta - \Delta_w f' - d\kappa,
\end{eqnarray}
where we have introduced the dimensionless temperature $u = (T_\infty - T)c_p/L$, the overheating $\Delta = (T_{\infty}-T_M)c_p/L$, and the measure for the strength of the short ranged forces, $\Delta_w = T_M \bar{\sigma} c_p/(L^2\delta)$, as well as the capillary length $d = T_M \sigma_{sl} c_p/L^2$.
As before, $T_\infty>T_M$ is the temperature far away from the melting front.

In the dimensionless form, the heat diffusion in the bulk is described by
\begin{equation}
\label{govEqs1}
D \nabla^2 u = \partial u/\partial t.
\end{equation}
The latent heat absorption at the moving interface is described by the Stefan condition
\begin{equation}
\label{govEqs2}
v_n = D \vec{n}\cdot \left(  \nabla u_L - \nabla u_S \right)|_{int},
\end{equation}
with the interface normal $\vec{n}$ and the normal component $v_n$ of the interface velocity.

The above equations can be combined to a single closed integro-differential equation in the steady state regime, as worked out in detail in \cite{Huter:2014uq},
\begin{eqnarray}
\label{BIRepresentation}
&&\Delta + \Delta_w \exp\left(\frac{- 2 |{y}| \rho}{\delta} \right)- \frac{d}{\rho}  \kappa  \\ \nonumber
&=&  \frac{p}{\pi} \int\limits_{-\infty}^{\infty} dy' \exp\left(-p[x(y) - x(y')]\right) K_0( p |\vec{r}-\vec{r}'|).
\end{eqnarray} 
with the integration along the solid-melt interface, which is parametrized by the dimensionless contour function $x(y)$.
$|\vec{r}-\vec{r}'|$ is the two-dimensional distance between the points of integration and observation.
All length scales are here rescaled by the tip radius of curvature $\rho$ of the asymptotic parabola far behind the triple junction, and $K_0$ is the modified Bessel function of second kind in zeroth order.
At the tip, the description is supplemented by the boundary condition of given value $\phi_0$.
From the tail region we get the classical Ivantsov relation between the overheating $\Delta$ and the Peclet number $p=\rho v/2D$ with the steady state velocity $v$, see \cite{Ivantsov},
\begin{equation} \label{iveq}
\Delta = \sqrt{\pi p}\, e^p \mathrm{erfc}(\sqrt{p}).
\end{equation}
This way, Eqs.~(\ref{dimLessLocEq})-(\ref{govEqs2}) are written in a closed form as a nonlinear eigenvalue problem to determine the interface contour $x(y)$ and the scale $d/\rho$, where the assumption has been made that the microscopic contact angle is small, $\phi_0\ll 1$. 
Such an integral equation approach is advantageous to treat the complex multi-scale problem accurately in a closed framework, similar to other moving boundary problems \cite{JSLangerLATurski_ActaMetall_25_1113_1977, CHueter_GBoussinot_EABrener_PRE_83_050601_2011, TFischaleckKKassner_EPL_81_54004_2008, Boussinot:2014fk}.

\begin{figure}
\begin{center}
\includegraphics[width=8cm]{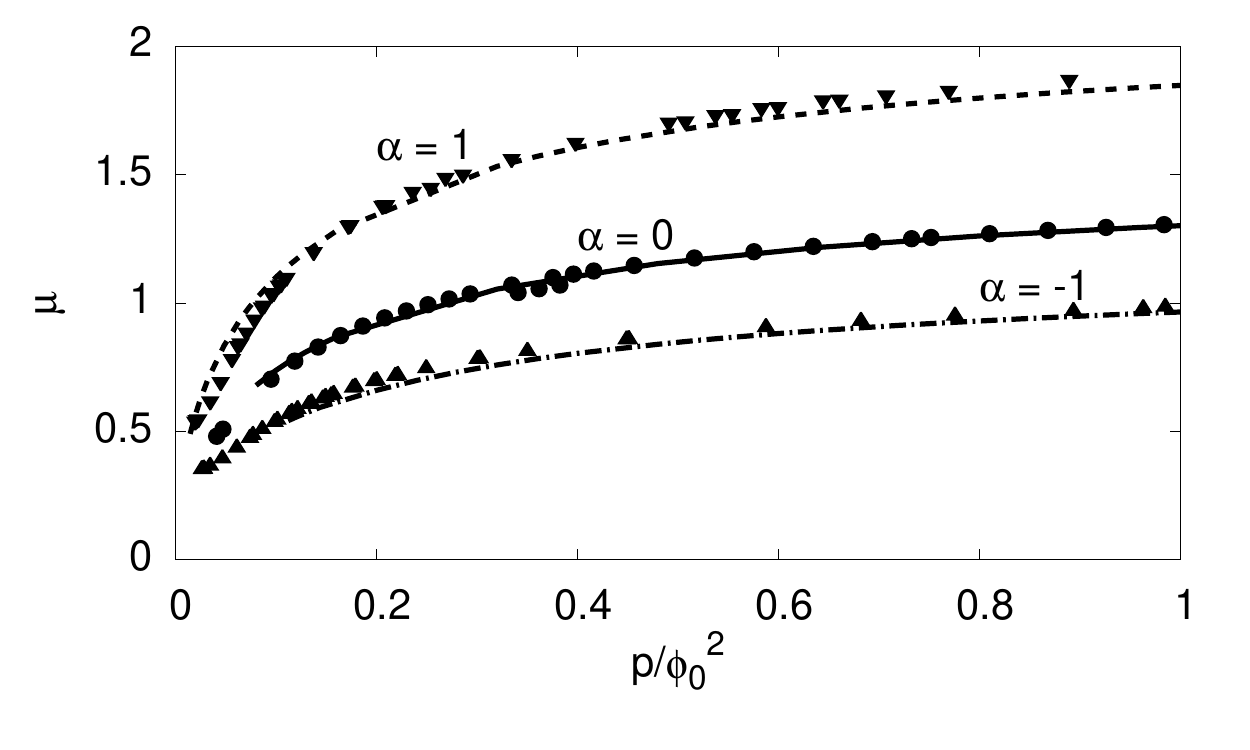}
\caption{\label{fullMu} 
The eigenvalue $\mu = d \phi_0^3/(\rho\Delta)$, which is a measure for the melting velocity, as function of $p/\phi_0^2$ for $\beta = 10$ and $\phi_0=0.025$.
The solid curves are obtained from the numerical solution of the full nonlinear steady state problem (\ref{BIRepresentation}), the symbols from the linearised description (\ref{linNumRepresentation}).
For low values of $p/\phi_0^2$, which is a measure for the overheating, we obtain a good agreement between the two approaches.
}
\end{center}
\end{figure}

\begin{figure}
\begin{center}
\includegraphics[width=8cm]{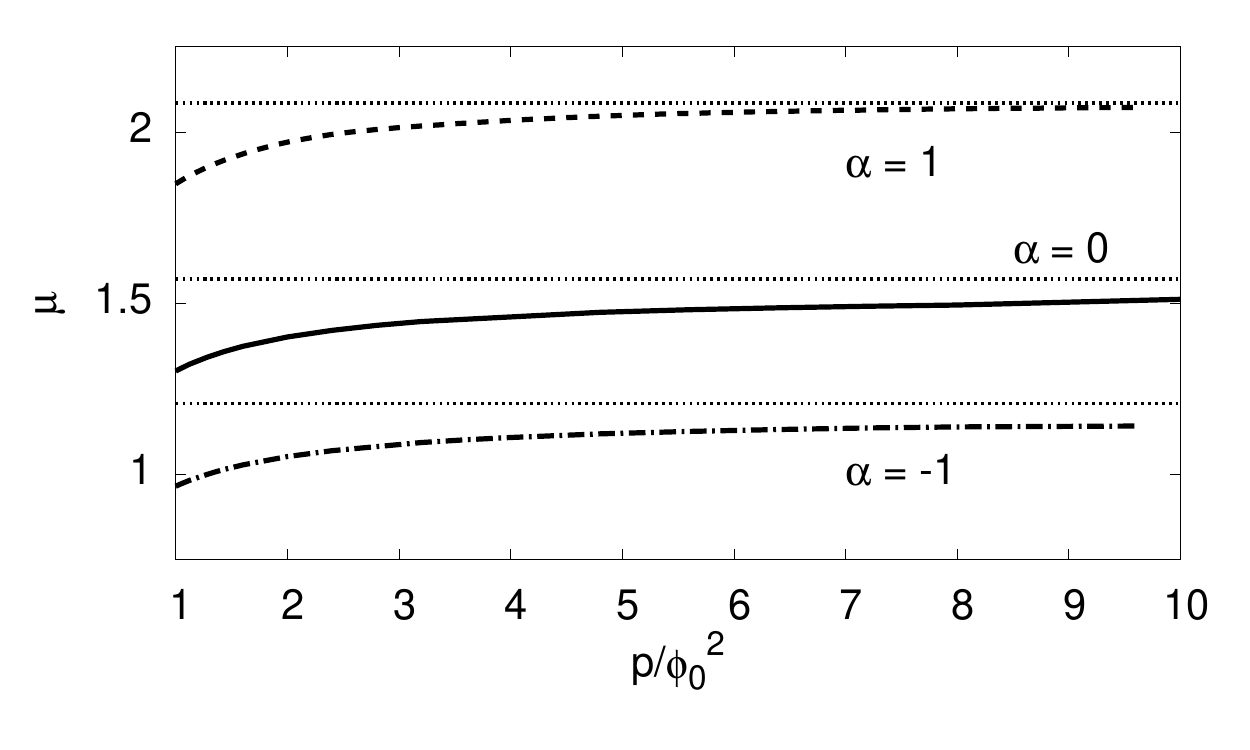}
\caption{\label{fullMu2} 
The same data as in Fig.~\ref{fullMu} in the regime of larger values $p/\phi_0^2$.
Here the results converge towards the analytical solutions of Eq.~(\ref{mu_cubeRoot}), which are shown as dotted horizontal lines. 
}
\end{center}
\end{figure}

First, from the full self-consistent solution of Eq.~(\ref{dimLessLocEq}), one can obtain the interface contours, which are exemplarily shown in Fig.~\ref{geometry}.
Moreover, also the eigenvalue $\mu = d \phi_0^3/(\rho\Delta)$, which is a measure for the lateral melting velocity, can be obtained.
The results are shown as continuous curves in Figs.~\ref{fullMu} and \ref{fullMu2} for two different regimes of the parameter $p/\phi_0^2$.
We note that this parameter effectively measures the overheating, as $p$ and $\Delta$ are directly connected via the Ivantsov relation (\ref{iveq}).
All other parameters can be combined to the quantities $\alpha = \Delta_w/\Delta$, which measures the strength of the short ranged interaction and $\beta = 2 d \phi_0^2/(\delta \Delta)$ for the range of the interaction.
We find in particular that attractive interactions $\alpha<0$ lead to a smaller eigenvalue $\mu$ and growth velocity, in agreement with the intuitive expectation.

The problem can be simplified in the regime of small opening angles $\phi_0\ll 1$ and small overheating, $\Delta\ll 1$.
This allows to linearise the curvature term and the integral kernel in Eq.~(\ref{BIRepresentation}), as well as to use the approximation $|dy/dx| \approx \phi_0$ in the disjoining potential.
As a result, one obtains a linear equations for the slope of the interface profile $y(x)$,
\begin{eqnarray}
&& 1 + \alpha \exp\left( -\frac{x \beta}{\mu}\right) + \mu \frac{d^2 y}{dx^2} = \frac{2p^{1/2}}{\pi^{3/2} \phi_0} \times \label{linNumRepresentation} \\
&\times&   \int\limits_0^{\infty} dx' \frac{dy}{dx'} \exp\left(-\frac{p(x' - x)}{\phi_0^2}\right) K_0 \left( \frac{p |x' - x|}{\phi_0^2} \right). \nonumber  
\end{eqnarray}
It is solved numerically, and the obtained eigenvalues $\mu$ are shown in Fig.~\ref{fullMu} as isolated points, which are in excellent agreement with the solution of the full problem.

For a further reduction of the parameter regime within the above limits $\Delta\ll 1$, $\phi_0\ll 1$, we can investigate the case $p/\phi_0^2 \gg 1$.
Then the integral equation (\ref{linNumRepresentation}) can be further simplified by using the large argument limit of the Bessel function and a truncation of the integral to the point of observation, since the contributions from outside this region are negligible.
It leads to the following equation for $dy/dx$,
\begin{eqnarray} \label{linEq}
&& 1 + \mu \frac{d^2y}{dx^2} + \alpha \exp\left(-\frac{\beta}{\mu}x\right) \nonumber \\
&=& \frac{\sqrt 2}{\pi} \int_{0}^{x} \frac{1}{(x - x')^{1/2}} \frac{dy}{dx'} dx',
\end{eqnarray}
which can be solved by Laplace transformation.
The eigenvalue $\mu$ can then be determined fully analytically.
\begin{equation} \label{mu_cubeRoot}
\mu =  \frac{\pi}{2}  \left[ \frac{1+\alpha-\beta}{2}  + \sqrt{\left(\frac{1+\alpha-\beta}{2} \right)^2 + \beta} \right]^3,
\end{equation}
See \cite{Huter:2014uq} for details.
Also here we find a very good agreement with the solution of the full problem, as shown in Fig.~\ref{fullMu2}.
Since $\mu\simeq d\phi_0^3 v\pi/(2D\Delta^3)$ for low overheating ($\Delta\simeq (\pi p)^{1/2}$ there), the velocity is proportional to the eigenvalue $\mu$, and allows to easily convert back the dimensionless parameters to observable quantities.

Finally, let us compare the predictions from this theory to the previous phase field simulations, see Fig.~\ref{figPF2}.
Assuming a microscopic angle $\phi_0\approx0.06$ and a relation between the range of the exponential decay of the analytical disjoining potential and the phase field predictions as $\delta = \xi/3$, we get for $\Delta=0.2$, $\alpha\approx-5.9$ and $\beta\approx 0.4$ in the phase field simulations and a corresponding ratio $p/\phi_0^2\approx 4.7$, which is reasonably well in the range of applicability of the analytical formula (\ref{mu_cubeRoot}).
With this, the measured velocity in the phase field model is close to the analytical prediction.
For a lower overheating, $\Delta=0.15$, the analytically calculated velocity is about $30\%$ lower than the phase field result.
Taking into account the uncertainty of the parameter $\phi_0$, the restrictions of the analytical expression (\ref{mu_cubeRoot}) and additional effects due to the finite interface thickness, the agreement is reasonable.
A doubling of the interface thickness, i.e.~of the range of the interaction, for $\Delta=0.2$ leads approximately to a doubling of the trijunction velocity in the phase field simulations (Fig.~\ref{fullMu2}), which agrees well with the analytical prediction (\ref{mu_cubeRoot}).
We can therefore conclude that the phase field and sharp interface predictions essentially coincide.
Further, more extended comparisons will be the subject of future investigations.

\section{Summary and conclusions}
\label{summary}

In this paper, we have analysed the multi-order parameter phase field model by Steinbach and Pezzolla concerning its ability to describe grain boundary premelting.
We find that in this model the transition between wetting and non-wetting states at the melting point is located at the classical threshold $\sigma_{gb}=2\sigma_{sl}$.
The range of the interaction is finite, as it is mediated by the overlap of the phase field interface profiles, which have a strictly finite width in this model.
The analytical predictions are quantitatively confirmed by numerical simulations using the {\sc OpenPhase} framework for repulsive interactions.

For attractive interactions, phase field simulations are used to predict the kinetics of grain boundary melting along an overheated dry grain boundary.
After an initial transient, a steady state regime is reached.
To further shed light on the role of short range interactions in a quantitative manner, a sharp interface model has been developed, which is solved by means of Greens function methods.
For small overheatings, microscopic tip angles $\phi_0$ and large values of $p/\phi_0^2$ the propagation velocity can be predicted analytically.
These predictions are in good agreement with the phase field simulations.

\section*{Acknowledgements}

This work has been supported by the Deutsche Forschungsgemeinschaft via the Priority Program SPP 1713 and the Max-Planck graduate school IMPRS-SurMat.
Support for the {\sc OpenPhase} framework by A. Monas is gratefully acknowledged.

\section*{References}

\bibliographystyle{elsarticle-num} 
\bibliography{references}

\begin{thebibliography}{10}
\expandafter\ifx\csname url\endcsname\relax
  \def\url#1{\texttt{#1}}\fi
\expandafter\ifx\csname urlprefix\endcsname\relax\def\urlprefix{URL }\fi
\expandafter\ifx\csname href\endcsname\relax
  \def\href#1#2{#2} \def\path#1{#1}\fi

\bibitem{Rappaz:2003aa}
M.~Rappaz, A.~Jacot, W.~J. Boettinger, Last-stage solidification of alloys:
  Theoretical model of dendrite-arm and grain coalescence, Metallurgical and
  Materials Transactions A 34 (2003) 467.

\bibitem{Mishin:2009aa}
Y.~Mishin, W.~J. Boettinger, J.~A. Warren, G.~B. McFadden, Thermodynamics of
  grain boundary premelting in alloys. {I. P}hase-field modeling, Acta
  Materialia 57 (2009) 3771.

\bibitem{Glicksman:1972aa}
M.~E. Glicksman, C.~L. Void, Heterophase dislocations - an approach towards
  interpreting high temperature grain boundary behavior, Surface Science 31
  (1972) 50.

\bibitem{Hsieh19891637}
T.~E. Hsieh, R.~W. Balluffi, Experimental study of grain boundary melting in
  aluminum, Acta Metallurgica 37 (1989) 1637.

\bibitem{Alsayed:2005aa}
A.~M. Alsayed, M.~F. Islam, J.~Zhang, P.~J. Collings, A.~G. Yodh, Premelting at
  defects within bulk colloidal crystals, Science 309 (2005) 1207.

\bibitem{Widom:1978aa}
B.~Widom, Structure of the $\alpha\gamma$ interface, J. Chem. Phys. 68 (1978)
  3878.

\bibitem{Besold:1994aa}
G.~Besold, O.~G. Mouritsen, Grain-boundary melting: A {M}onte {C}arlo study,
  Phys. Rev. B 50 (1994) 6573.

\bibitem{PhysRevB.21.1893}
R.~Kikuchi, J.~W. Cahn, Grain-boundary melting transition in a two-dimensional
  lattice-gas model, Phys. Rev. B 21 (1980) 1893.

\bibitem{PhysRevE.79.020601}
J.~J. Hoyt, D.~Olmsted, S.~Jindal, M.~Asta, A.~Karma, Method for computing
  short-range forces between solid-liquid interfaces driving grain boundary
  premelting, Phys. Rev. E 79 (2009) 020601.

\bibitem{Williams20093786}
P.~L. Williams, Y.~Mishin, Thermodynamics of grain boundary premelting in
  alloys. {II. A}tomistic simulation, Acta Materialia 57~(13) (2009) 3786.

\bibitem{PhysRevE.81.051601}
N.~Wang, R.~Spatschek, A.~Karma, Multi-phase-field analysis of short-range
  forces between diffuse interfaces, Phys. Rev. E 81 (2010) 051601.

\bibitem{Tang:2006aa}
M.~Tang, W.~C. Carter, R.~M. Cannon, Diffuse interface model for structural
  transitions of grain boundaries, Phys. Rev. B. 73 (2006) 024102.

\bibitem{Lobkovsky2002202}
A.~E. Lobkovsky, J.~A. Warren, Phase field model of premelting of grain
  boundaries, Physica D: Nonlinear Phenomena 164 (2002) 202.

\bibitem{PhysRevB.78.184110}
J.~Mellenthin, A.~Karma, M.~Plapp, Phase-field crystal study of grain-boundary
  premelting, Phys. Rev. B 78 (2008) 184110.

\bibitem{Adland:2013ys}
A.~Adland, A.~Karma, R.~Spatschek, D.~Buta, M.~Asta, Phase-field-crystal study
  of grain boundary premelting and shearing in bcc iron, Phys. Rev. B 87 (2013)
  024110.

\bibitem{PhysRevB.77.224114}
J.~Berry, K.~R. Elder, M.~Grant, Melting at dislocations and grain boundaries:
  A phase field crystal study, Phys. Rev. B 77 (2008) 224114.

\bibitem{Olmsted:2011vn}
D.~L. Olmsted, D.~Buta, A.~Adland, S.~M. Foiles, M.~Asta, A.~Karma,
  Dislocation-pairing transitions in hot grain boundaries, Phys. Rev. Lett. 106
  (2011) 046101.

\bibitem{Spatschek:2010fk}
R.~Spatschek, A.~Karma, Amplitude equations for polycrystalline materials with
  interaction between composition and stress, Phys. Rev. B 81 (2010) 214201.

\bibitem{Huter:2014aa}
C.~H\"uter, C.-D. Nguyen, R.~Spatschek, J.~Neugebauer, Scale bridging between
  atomistic and mesoscale modelling: applications of amplitude equations
  descriptions, Modelling Simul. Mater. Sci Eng. 22 (2014) 034001.

\bibitem{kar13}
R.~Spatschek, A.~Adland, A.~Karma, Structural short-range forces between
  solid-melt interfaces, Physical Review B 87 (2013) 024109.

\bibitem{KarmaReview}
A.~Karma, Phase-field methods, in: K.~Buschow, et~al. (Eds.), Encyclopedia of
  Materials Science and Technology, Elsevier, Oxford, 2001, p. 6873.

\bibitem{ChenReview}
L.~Q. Chen, Phase-field models for microstructure evolution, Annu. Rev. Mater.
  Res. 32 (2002) 113.

\bibitem{steinbach}
I.~Steinbach, Phase-field models in materials science, Modelling Simul. Mater.
  Sci. Eng. 17 (2009) 073001.

\bibitem{Steinbach:2013aa}
I.~Steinbach, Phase-field model for microstructure evolution at the mesoscopic
  scale, Annu. Rev. Mater. Res. 43 (2013) 89.

\bibitem{Spatschek11}
R.~Spatschek, E.~Brener, A.~Karma, Phase field modeling of crack propagation,
  Phil. Mag. 91 (2011) 75.

\bibitem{Steinbach:1998aa}
I.~Steinbach, F.~Pezzolla, A generalized field method for multiphase
  transformations using interface fields, Physica D 115 (1998) 87.

\bibitem{EikenJ2006}
J.~Eiken, B.~B{\"o}ttger, I.~Steinbach, Multiphase-field approach for
  multicomponent alloys with extrapolation scheme for numerical application,
  Phys. Rev. E 73 (2006) 066122.

\bibitem{openphase}
\href{www.openphase.de}{www.openphase.de} [online].

\bibitem{Folch:2005kx}
R.~Folch, M.~Plapp, Quantitative phase-field modeling of two-phase growth,
  Phys. Rev. E. 72 (2005) 011602.

\bibitem{Bhogireddy:2014aa}
V.~S. P.~K. Bhogireddy, C.~H\"uter, J.~Neugebauer, I.~Steinbach, A.~Karma,
  R.~Spatschek, Phase-field modeling of grain-boundary premelting using
  obstacle potentials, Phys. Rev. E 90 (2014) 012401.

\bibitem{Frolov:2011fk}
T.~Frolov, Y.~Mishin, Liquid nucleation at superheated grain boundaries, Phys.
  Rev. Lett. 106 (2011) 155702.

\bibitem{Huter:2014uq}
C.~H\"uter, F.~Twiste, E.~A. Brener, J.~Neugebauer, R.~Spatschek, The influence
  of short range forces on melting along grain boundaries, Phys. Rev. B 89
  (2014) 224104.

\bibitem{EAB_CH_DP_DET_PRL99_105701_2007}
E.~A. Brener, C.~H\"uter, D.~Pilipenko, D.~E. Temkin, Velocity selection
  problem in the presence of triple junction, Physical Review Letters 99 (2007)
  105701.

\bibitem{Read:1950fk}
W.~T. Read, W.~Shockley, Dislocation models of crystal grain boundaries, Phys.
  Rev. 78 (1950) 275.

\bibitem{openmp}
\href{http://openmp.org}{http://openmp.org} [online].

\bibitem{Guo:2011aa}
W.~Guo, R.~Spatschek, I.~Steinbach, An analytical study of the static state of
  multi-junctions in a multi-phase field model, Physica D 240 (2011) 382.

\bibitem{Ivantsov}
G.~P. Ivantsov, The temperature field around a spherical, cylindrical or
  pointed crystal growing in a cooling solution, Dokl. Akad. Nauk USSR 58
  (1947) 567--569.

\bibitem{JSLangerLATurski_ActaMetall_25_1113_1977}
J.~S. Langer, L.~A. Turski, Studies in the theory of interfacial stability-i.
  stationary symmetric model, Acta Metallurgica 25 (1976) 1113.

\bibitem{CHueter_GBoussinot_EABrener_PRE_83_050601_2011}
G.~Boussinot, C.~H\"uter, E.~A. Brener, D.~E. Temkin, Growth of two phase
  finger in eutectic systems, Phys. Rev. E 83 (2011) 050601.

\bibitem{TFischaleckKKassner_EPL_81_54004_2008}
T.~Fischaleck, K.~Kassner, Extending the scope of microscopic solvability :
  Combination of the kruskal-segur method with zauderer decomposition, Euro.
  Phys. Lett. 81 (2008) 54004.

\bibitem{Boussinot:2014fk}
G.~Boussinot, C.~H\"uter, R.~Spatschek, E.~A. Brener, Isothermal solidification
  in peritectic systems, Acta Materialia 75 (2014) 212.

\end{thebibliography}

\end{document}